\documentclass[aps, pra, 10pt, twocolumn, superscriptaddress, floatfix, longbibliography]{revtex4-1}
\usepackage{anyfontsize}
\usepackage{graphicx}
\usepackage{placeins} % float handling
\usepackage{amssymb}
\usepackage{amsmath}
\usepackage{braket}
\usepackage{bm}
\usepackage{verbatim}
\usepackage{enumitem}
\usepackage[unicode=true, bookmarks=false, breaklinks=false,pdfborder={0 0 1},backref=false,colorlinks=true]{hyperref}
\hypersetup{
 linkcolor=[rgb]{0,0,1},citecolor=[rgb]{0,0,1},urlcolor=[rgb]{0,0,1}}
\mathchardef\myphen="2D
\begin{document}
\title{Two-dopant origin of competing stripe and pair formation in Hubbard and \textit{t-J} models}
\author{Tizian~Blatz}
\email{blatz.tizian@physik.uni-muenchen.de}
\affiliation{Department of Physics and Arnold Sommerfeld Center for Theoretical Physics (ASC), Ludwig-Maximilians-Universit\"at M\"unchen, Theresienstr. 37, M\"unchen D-80333, Germany}
\affiliation{Munich Center for Quantum Science and Technology (MCQST), Schellingstr. 4, M\"unchen D-80799, Germany}
\author{Ulrich~Schollwöck}
\affiliation{Department of Physics and Arnold Sommerfeld Center for Theoretical Physics (ASC), Ludwig-Maximilians-Universit\"at M\"unchen, Theresienstr. 37, M\"unchen D-80333, Germany}
\affiliation{Munich Center for Quantum Science and Technology (MCQST), Schellingstr. 4, M\"unchen D-80799, Germany}
\author{Fabian~Grusdt}
\affiliation{Department of Physics and Arnold Sommerfeld Center for Theoretical Physics (ASC), Ludwig-Maximilians-Universit\"at M\"unchen, Theresienstr. 37, M\"unchen D-80333, Germany}
\affiliation{Munich Center for Quantum Science and Technology (MCQST), Schellingstr. 4, M\"unchen D-80799, Germany}
\author{Annabelle~Bohrdt}
\email{a.bohrdt@lmu.de}
\affiliation{Munich Center for Quantum Science and Technology (MCQST), Schellingstr. 4, M\"unchen D-80799, Germany}
\affiliation{Institute of Theoretical Physics, University of Regensburg, Regensburg D-93053, Germany}
%
%\date{\today}
\date{March 31, 2025}
\begin{abstract}
Understanding the physics of the two-dimensional Hubbard model is widely believed to be a key step in achieving a full understanding of high-$T_\mathrm{c}$ cuprate superconductors. In recent years, progress has been made by large-scale numerical simulations at finite doping and, on the other hand, by microscopic theories able to capture the physics of individual charge carriers.
In this work, we study single pairs of dopants in a cylindrical system using the density-matrix renormalization group algorithm.
We identify two coexisting charge configurations that couple to the spin environment in different ways:
A tightly bound configuration featuring (next-)nearest-neighbor pairs and a stripe-like configuration of dopants on opposite sides of the cylinder, accompanied by a spin domain wall.
Thus, we establish that the interplay between stripe order and uniform pairing, central to the models' phases at finite doping, has its origin at the single-pair level.
By interpolating between the Hubbard and the related $t$-$J$ model, we are able to quantitatively understand discrepancies in the pairing properties of the two models
through the three-site hopping term usually omitted from the $t$-$J$ Hamiltonian.
This term is closely related to a next-nearest-neighbor tunneling $t'$, which we observe to upset the balance between the competing stripe and pair states on the two-dopant level.
\end{abstract}
\maketitle
%
%%%%%%%%%%%%%%%%%%%%%%%%%%%%%%%%%%
\section{Introduction}
\begin{figure}[t!!]
\centering
\includegraphics[width=\linewidth]{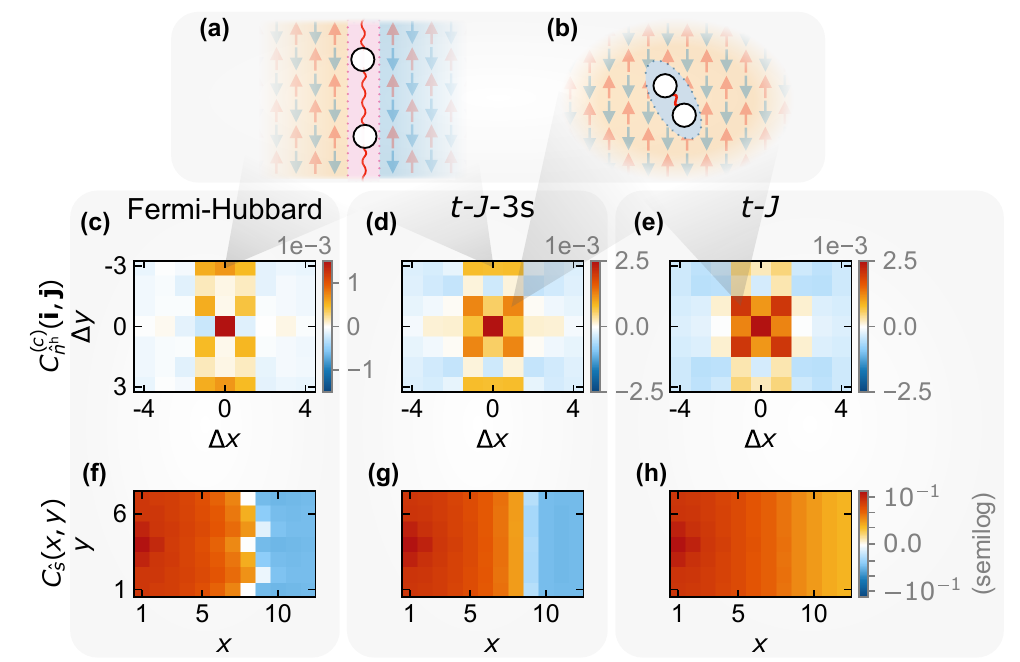}
%\caption{
\caption{
{Pair Structure:}~Charge and spin correlations for a pair of dopants in the Fermi-Hubbard, $t$-$J$-3s, and $t$-$J$ models on $6$-leg cylinders.
\textbf{(a, b)} We identify a tightly bound and a stripe-like configuration of the holes: In the stripe configuration (a) the holes reside on opposite sides of the cylindrical system, accompanied by a spin domain wall. In the tightly bound configuration (b), the holes form a (next-)nearest-neighbor pair in a uniform AFM background.
\textbf{(c, d, e)} Connected density-density correlation function relative to a reference position, averaged over the center of the lattice.
The color scale cuts off the strong autocorrelation at $(0, 0)$.
For the Fermi-Hubbard model, the correlations are corrected for doublon-hole fluctuations.
\textbf{(f, g, h)} Staggered spin-spin correlation function, relative to a reference position at the edge of the cylinder.
Positive values indicate uninterrupted AFM order in the $t$-$J$ model, while a domain wall in the AFM manifests as a sign change in the Fermi-Hubbard and $t$-$J$-3s models.
}
\label{fig_overview_correlators}
\end{figure}

Engineering superconducting materials with improved properties will most likely require a microscopic understanding of unconventional superconductors such as the cuprates~\cite{Bednorz1986-PossibleHighTcSuperconductivity, Kastner1998-MagneticTransportOptical, Scalapino2012-CommonThreadPairing}.
Despite recent progress~\cite{Qin2020-AbsenceSuperconductivityPureb, Jiang2021-GroundstatePhaseDiagram, Jiang2024-GroundstatePhaseDiagram, Lu2024-EmergentSuperconductivityCompeting, Shen2024-GroundStateElectrondopeda, Xu2024-CoexistenceSuperconductivityPartially} in numerically determining the ground states of the two-dimensional single-band toy models believed to contain the relevant physics~\cite{Anderson1987-ResonatingValenceBonda, Zhang1988-EffectiveHamiltonianSuperconducting}, we still lack a theoretical framework that would allow efficient predictions guiding the search for new materials.
While studies of the Fermi-Hubbard model~\cite{Hubbard1963-ElectronCorrelationsNarrowb} report superconducting domes on both the electron and hole doped side~\cite{Xu2024-CoexistenceSuperconductivityPartially}, weaker or even no superconductivity is found~\cite{Jiang2021-GroundstatePhaseDiagram, Jiang2022-PairingPropertiesText, Lu2024-EmergentSuperconductivityCompeting} on the hole-doped side in the closely related $t$-$J$ model~\cite{Chao1977-KineticExchangeInteraction, Chao1978-CanonicalPerturbationExpansion, Hirsch1985-AttractiveInteractionPairing}. \\
As we highlight in this work, an important ingredient to resolving this puzzle is the three-site (or singlet) hopping term arising from the Schrieffer-Wolff transformation connecting the $t$-$J$ and Fermi-Hubbard models. This term is usually not included in studies of the $t$-$J$ model but could be vital to the pairing properties as it has been argued to mediate a hole-hole repulsion~\cite{Ammon1995-EffectThreesiteHopping, Coulthard2018-GroundstatePhaseDiagram} and found to remedy key discrepancies in the single-particle spectral function~\cite{Kuzmin2014-ComparisonElectronicStructure, Wang2015-OriginStrongDispersion}.
As it allows next-nearest-neighbor hopping processes for dopant charge carriers, its effects are intertwined with those of the respective $t'$ tunneling term which has already been established to be crucial for superconductivity~\cite{Qin2020-AbsenceSuperconductivityPureb, Xu2024-CoexistenceSuperconductivityPartially}. \\
The different models and interaction parameters are accompanied by a sizable array of competing and coexisting orderings, including
a Mott-insulating state featuring antiferromagnetic (AFM) correlations,
uniform-density d-wave superconductivity,
stripes -- which we identify with a charge-density wave accompanied by spin domain walls --
and various crystalline phases (e.g. denoted as WC* or W3 in the literature~\cite{Jiang2021-GroundstatePhaseDiagram, Jiang2024-GroundstatePhaseDiagram}).
\\
Two central building blocks for a microscopic understanding of these phases are single dopants -- known as magnetic polarons -- and pairs of dopants in an AFM background.
The microscopic pairing mechanism in particular has recently gained renewed interest and has been linked to the Hamiltonian's sign structure~\cite{Lu2024-SignStructure$ttextensurematht^ensuremathtextensuremathJ$, Chen2023-DWavePairDensityWaveSuperconductivitya}.
With the advent of quantum-simulation experiments~\cite{Bloch2012-QuantumSimulationsUltracold, Gross2017-QuantumSimulationsUltracold} affording single-atom, single-site resolution in extended lattices, significant progress has been made~\cite{Bohrdt2021-ExplorationDopedQuantuma}, in particular concerning  polarons~\cite{Chiu2019-StringPatternsDoped, Koepsell2021-MicroscopicEvolutionDopeda, Ji2021-CouplingMobileHolea, Lebrat2024-ObservationNagaokaPolarons, Prichard2024-DirectlyImagingSpin}, and most recently pairing~\cite{Hirthe2023-MagneticallyMediatedHolea} and stripe formation~\cite{Bourgund2025-FormationIndividualStripes}.
\\
In this work, we investigate a single pair of dopants to determine which of the finite-doping puzzles can be traced back to this minimial building block. We use the density-matrix renormalization group (DMRG)~\cite{White1992-DensityMatrixFormulationb, White1993-DensitymatrixAlgorithmsQuantuma} algorithm
to extract the ground state properties of cylindrical systems of width $6$. In both the Fermi-Hubbard and $t$-$J$ models, we find bound pairs of dopants in a superposition of a tightly bound configuration and a single $1/3$ filled stripe, as sketched in Fig.~\ref{fig_overview_correlators}a;b.
By interpolating between the Fermi-Hubbard and $t$-$J$ Hamiltonians, we can quantitatively trace back discrepancies in the relative weights of the two charge contributions and the formation of a spin domain wall in the Fermi-Hubbard model to the three-site hopping term. We probe the interplay between charge and magnetic order and conclude by giving an outlook on the related $t'$ term.
Overall, our work establishes a competition of two charge configurations of a single pair of dopants as the likely microscopic origin of more complex phases found at finite doping.
%%%%%%%%%%%%%%%%%%%%%%%%%%%%%%%%%%
\section{Models}
At the core of our work, we consider the Fermi-Hubbard model
\begin{align}
\begin{split}
    \hat{H}_\mathrm{FH} = \
    &- t \sum_{\langle \mathbf{i}, \mathbf{j} \rangle, \sigma} (\hat{c}^\dagger_{\mathbf{i}, \sigma} \hat{c}_{\mathbf{j}, \sigma} + \mathrm{H.c.})
    + U \sum_\mathbf{i} \hat{n}_{\mathbf{i} \uparrow} \hat{n}_{\mathbf{i} \downarrow}
\end{split}
\label{eq:Fermi-Hubbard}
\end{align}
characterized by the tunneling $t$ and on-site interaction $U$. Here, $\hat{c}^{(\dagger)}_{\mathbf{i}, \sigma}$ is the fermionic annihilation (creation) operator at coordinate $\mathbf{i} = (x, y)$ with spin $\sigma$ and $\hat{n}_\mathbf{i} = \sum_\sigma \hat{n}_{\mathbf{i}, \sigma} = \sum_\sigma \hat{c}^{\dagger}_{\mathbf{i}, \sigma} \hat{c}_{\mathbf{i}, \sigma}$ is the particle number operator.
In the strong-coupling limit $t/U \ll 1$, the $t$-$J$-3s model
\begin{align}
\begin{split}
    \hat{H}_{t \myphen J \myphen 3 \mathrm{s}} = \
    &- t \sum_{\langle \mathbf{i}, \mathbf{j} \rangle, \sigma} (\Tilde{c}^\dagger_{\mathbf{i}, \sigma} \Tilde{c}_{\mathbf{j}, \sigma} + \mathrm{H.c.}) \\
    &+ J \sum_{\langle \mathbf{i}, \mathbf{j} \rangle} \left( \mathbf{\hat{S}}_{\mathbf{i}} \cdot \mathbf{\hat{S}}_{\mathbf{j}}
    - \frac{\Tilde{n}_i \Tilde{n}_j}{4} \right) \\
    &- t_3 \sum_{\langle \mathbf{i}, \mathbf{j}, \mathbf{k} \rangle}
    (\Tilde{b}^\dagger_{\mathbf{i}, \mathbf{j}} \Tilde{b}_{\mathbf{j}, \mathbf{k}}
    + \mathrm{H.c.})
\end{split}
\label{eq:t-J}
\end{align}
emerges as the lowest-order approximation in $t/U$ from the Schrieffer-Wolff transformation~\cite{Auerbach1994-InteractingElectronsQuantuma}.
Double occupancies are eliminated, and the creation operators are replaced by $\Tilde{c}^\dagger_{\mathbf{i}, \sigma} = \hat{c}^\dagger_{\mathbf{i}, \sigma} ( 1 - \hat{n}_{\mathbf{i}, - \sigma})$.
From $\Tilde{c}^{(\dagger)}_{\mathbf{i}, \sigma}$, we define the corresponding number operator $\Tilde{n}_{\mathbf{i}, \sigma}$ and the singlet annihilation operator
\begin{align}
    \Tilde{b}_{\mathbf{i}, \mathbf{j}}
    = \frac{1}{\sqrt{2}} (\Tilde{c}_{\mathbf{i}, \downarrow} \Tilde{c}_{\mathbf{j}, \uparrow} - \Tilde{c}_{\mathbf{i}, \uparrow} \Tilde{c}_{\mathbf{j}, \downarrow}) \ .
\end{align}
The lowest-order terms in $t/U$ correspond to virtual hopping processes and give rise to the effective Heisenberg superexchange interaction with $J = 4 t^2 / U$ and the singlet (or three-site) hopping term with $t_3 = J / 2$.
Here, $\mathbf{\hat{S}}_\mathbf{i}$ denotes the Heisenberg spin operator at site $\mathbf{i}$ and $\langle \mathbf{i}, \mathbf{j}, \mathbf{k} \rangle$ restricts $\mathbf{i}$ and $\mathbf{k}$ to nearest neighbors of $\mathbf{j}$ with $\mathbf{i} \neq \mathbf{k}$.
Following the common convention, we call $\hat{H}_{t \myphen J} = \hat{H}_{t \myphen J \myphen 3 \mathrm{s}}(t_3 = 0)$ the $t$-$J$ model, omitting the singlet-hopping term.% ($t_3 = 0$).
Throughout this work, we consider $U/t = 8$, which fixes $J/t = 1/2$ and $t_3/t = 1/4$.
This value of $U$ is realistic for the study of cuprate materials~\cite{Hybertsen1989-CalculationCoulombinteractionParameters, Dagotto1994-CorrelatedElectronsHightemperature} and achievable in typical ultracold-atom setups~\cite{Mazurenko2017-ColdatomFermiHubbardb, Gall2021-CompetingMagneticOrders}. \\
For an in-depth analysis contrasting the models, we perform an interpolation controlled by the dimensionless parameter $\lambda \in [ 0, \ 1 ]$.
\begin{figure*}[t!!]
\centering
\includegraphics[width=\textwidth]{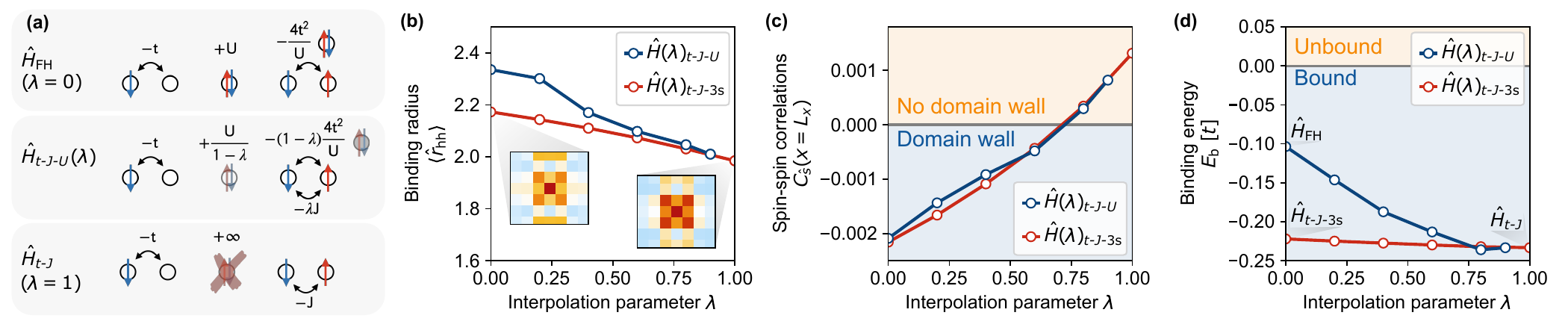}
\caption{
{Interpolation between Hamiltonians:}~\textbf{(a)} Illustration of the scan between microscopic models: Starting from $\hat{H}_\mathrm{FH}$ governed by $t$ and $U$, the scan parameter $\lambda$ gradually suppresses double occupancies and introduces explicit Heisenberg interactions $J$ to keep the combined magnetic interactions constant. At $\lambda = 1$ we arrive at the $t$-$J$ model.
\textbf{(b)} Hole-hole binding radius $\langle \hat{r}^\mathrm{hh} \rangle$ observed to decrease continuously with increasing $\lambda$, corresponding to the holes forming a more tightly bound pair in the $t$-$J$ model.
The data for $\hat{H}(\lambda)_{t \myphen J \myphen U}$ is corrected for doublon-hole fluctuations.
The insets show the corresponding 2D charge structures presented in Fig.~\ref{fig_overview_correlators}d,e.
\textbf{(c)} Staggered spin-spin correlations from one edge of the cylinder ($x = 1$) to the other ($x = L_x$).
For the Fermi-Hubbard and $t$-$J$-3s models at $\lambda=0$, a domain wall is observed which disappears between $\lambda=0.6$ and $\lambda=0.8$.
At $\lambda=1$, the $t$-$J$ model shows uninterrupted AFM order across the length of the cylinder.
\textbf{(d)} Binding energy $E_\mathrm{b}$ of the pair of doped holes as a function of $\lambda$.
Across the scan, the pair stays bound, with the binding energy nearly constant between the $t$-$J$-type models. The binding energy in the Fermi-Hubbard model is lower by more than a factor of $2$.
}
\label{fig_lambda_scan}
\end{figure*}
To interpolate between the Fermi-Hubbard and $t$-$J$ models, we consider the Hamiltonian
\begin{align}
\begin{split}
    \hat{H}_{t \myphen J \myphen U}(\lambda) = - t \: \hat{H}_t + \frac{U}{1 - \lambda} \: \hat{H}_U + \lambda J \: \hat{H}_J \ ,
\end{split}
\end{align}
where $\hat{H}_t$, $\hat{H}_U$ and $\hat{H}_J$ refer to the terms proportional to the respective parameters $t$ and $U$ in (\ref{eq:Fermi-Hubbard}), and $J$ in (\ref{eq:t-J}).
In this special case of a $t$-$J$-$U$ model~\cite{Spalek2022-SuperconductivityHighRelated}, the dependency of $\hat{H}_{t \myphen J \myphen U}(\lambda)$ on $\lambda$ is chosen such that the effective magnetic interaction
\begin{align}
    J_\mathrm{eff}
    = (1 - \lambda) \frac{4t^2}{U} + \lambda J
    = J
\end{align}
is independent of $\lambda$. The model realizes $\hat{H}_{t \myphen J \myphen U}(0) = \hat{H}_\mathrm{FH}$, and $\hat{H}_{t \myphen J \myphen U}(1) = \hat{H}_{t \myphen J}$.
To relate discrepancies between $\hat{H}_\mathrm{FH}$ and $\hat{H}_{t \myphen J}$ to the changes introduced by the singlet hopping term, we define a corresponding model interpolating between the $t$-$J$-3s and $t$-$J$ models as
\begin{align}
\begin{split}
    \hat{H}_{t \myphen J \myphen 3 \mathrm{s}}(\lambda)
    = - t \: \hat{H}_\mathrm{t} + J \: \hat{H}_\mathrm{J} - (1 - \lambda) t_3 \: \hat{H}_{t_3} \ ,
\end{split}
\end{align}
which again satisfies $\hat{H}_{t \myphen J \myphen 3 \mathrm{s}}(1) = \hat{H}_{t \myphen J}$. \\ 
In the main part of our work, we analyze the pairing properties of dopants across these models.
Due to particle-hole symmetry, our results do not distinguish between electron and hole dopants.
However, from this point onwards, we refer to the dopants as holes to facilitate a clear distinction between dopant holes and doublon-hole fluctuations appearing in the Fermi-Hubbard Hilbert space.
We introduce the hole number operator $\hat{n}^\mathrm{h}_\mathbf{i} = (1 - \hat{n}_{\mathbf{i}, \uparrow})(1 - \hat{n}_{\mathbf{i}, \downarrow})$, and define $N^\mathrm{h} = \langle \hat{N}^\mathrm{h} \rangle = \sum_\mathbf{i} \langle \hat{n}^\mathrm{h}_\mathbf{i} \rangle$. Ground-state searches are performed using DMRG on systems of size $L_y \times L_x = 6 \times 12$ with periodic boundary conditions in the y-direction and open boundary conditions in the x-direction.
The system is doped with $0$, $1$, or $2$ holes away from half-filling.
%%%%%%%%%%%%%%%%%%%%%%%%%%%%%%%%%%
\section{Hole Pairs}
We now turn to investigate the pairing of dopants in the Fermi-Hubbard, $t$-$J$-3s, and $t$-$J$ models. The central questions we address are:
\begin{enumerate}[label=\roman*)]
    \item Do the two dopants form a bound pair? If so, what is the pair's charge structure, and how does it relate to pairing and stripe formation at finite doping?
    \item How do these properties change between the models under investigation? How much of the difference between the FH and $t$-$J$ models is accounted for by the singlet hopping term appearing in $\hat{H}_{t \myphen J \myphen 3 \mathrm{s}}$?
    \item How are the spin and charge sectors connected?
\end{enumerate}
As a first step to answering these questions and to obtain insights into the charge and magnetic order, we consider the density-density and spin-spin correlation functions of the two-hole ground states across the $3$ models.
We define the connected density-density correlator for the holes as
\begin{align}
    C^{(c)}_{\hat{n}^\mathrm{h}}(\mathbf{i}, \mathbf{j}) = 
    \langle \hat{n}^\mathrm{h}_\mathbf{i} \hat{n}^\mathrm{h}_\mathbf{j} \rangle
    - \frac{\langle \hat{N}^\mathrm{h} \rangle - 1}{\langle \hat{N}^\mathrm{h} \rangle} \langle \hat{n}^\mathrm{h}_\mathbf{i} \rangle \langle \hat{n}^\mathrm{h}_\mathbf{j} \rangle \ ,
\label{eq:connected_corr}
\end{align}
where the normalization factor accounts for the finite number of holes in the system.
For each model, Fig.~\ref{fig_overview_correlators} shows the correlations between the site $\mathbf{j} = \mathbf{i_0} + (\Delta x, \Delta y)$ with respect to a reference site $\mathbf{i_0}$ averaged over the center of the system: $x_\mathbf{i_0} \in \left[ L_x / 2, L_x / 2 + 1 \right]$; $1 \leq y_\mathbf{i_0} \leq L_y$. \\
\\
In all three models, the density-density correlations offer a clear picture, indicative of real-space pairing -- with the dopant holes mainly occupying the same or adjacent rungs of the cylinder. \\
Comparing the $t$-$J$ and $t$-$J$-3s models, we identify a shift in weight between two main contributions to the correlation function:
In the $t$-$J$ model, the holes form a tightly bound pair with the correlation function assuming its largest value on next-nearest-neighbor sites, avoiding a kinetic-energy penalty encountered on nearest-neighbor sites~\cite{White1997-HolePairStructuresa}. \\
In comparison, the $t$-$J$-3s model features enhanced correlations around the position $(\Delta x, \Delta y) = (0, L_y / 2)$ relative to the reference hole, i.e., at the opposite side going around the cylinder.
\begin{figure*}[t!!]
\centering
\includegraphics[width=\textwidth]{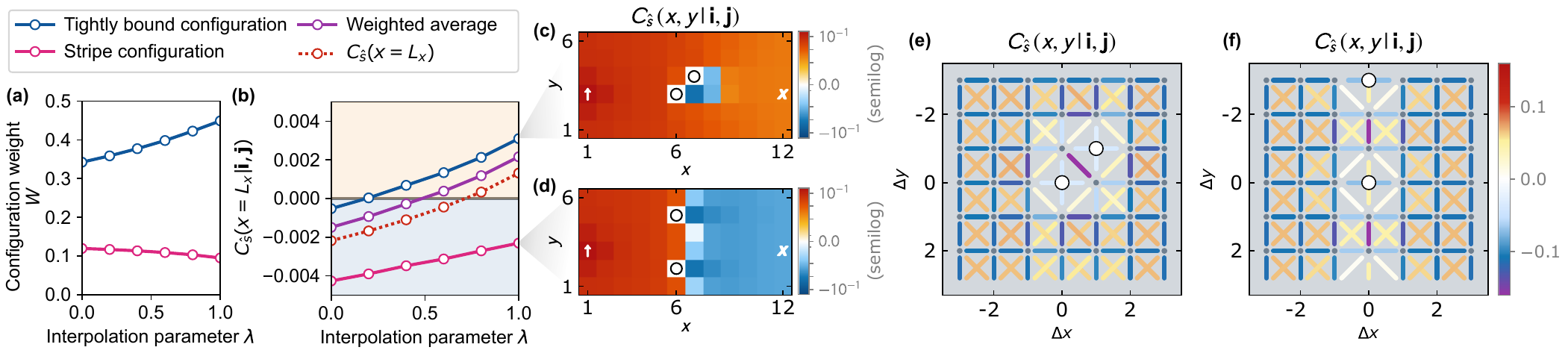}
%\caption{{
\caption{
{Pair Configurations:}~Quantitative comparison of the tightly bound configuration and the stripe configuration, interpolating between the $t$-$J$-3s and $t$-$J$ models.
\textbf{(a)} Weights $W$ of the specified hole configurations as defined in the main text via the hole-hole correlation function.
With increasing scan-parameter $\lambda$, the weight of the tightly bound configuration increases, while that of the stripe configuration decreases -- corresponding to the decrease in binding radius observed in Fig.~\ref{fig_lambda_scan}b.
\textbf{(b)} Spin-spin correlations across the lattice for the different hole configurations accessed via the 4-point correlation function $C_{\hat{s}}(x, y | \mathbf{i}, \mathbf{j})$, with $\mathbf{i}$ fixed in the center of the system.
For all values of $\lambda$, the stripe configuration ($\mathbf{i} - \mathbf{j} = (0, 3)$) features a domain wall, while the tightly bound configuration ($\mathbf{i} - \mathbf{j} = (1, 1)$) features uninterrupted AFM order for $\lambda \geq 0.2$.
The weighted average of the two contributions (according to the weights in (a)) explains the emergence of a domain wall in the full system presented in Fig.~\ref{fig_lambda_scan}c (dashed line showing $C_{\hat{s}}(x=L_x)$).
\textbf{(c, d)} $C_{\hat{s}}(x, y | \mathbf{i}, \mathbf{j})$ for the $t$-$J$ model with fixed hole positions of the holes (black circles) and the reference spin (white arrow).
This representation confirms uninterrupted AFM order for the tightly-bound configuration (c) and shows that the spin domain wall spatially coincides with the stripe-like configuration (d).
The white crosses mark the values of the correlation function across the system used for the respective lines in (b).
\textbf{(e, f)} Local spin-spin correlations around the respective charge configurations.
The tightly bound configuration on next-nearest-neighbor sites shares a $2 \times 2$ plaquette with two particles forming a singlet (c).
The stripe configuration is accompanied by spin singlets on the rung (d).
}
\label{fig_configuration_weights}
\end{figure*}
In conjunction with the appearance of a domain wall in the AFM pattern, we interpret this contribution as the $2$ holes forming a single $1/3$ filled stripe around the cylinder. \\
In the Fermi-Hubbard model, we unveil a similar pair structure as in the $t$-$J$-3s model, when correcting the correlation function for doublon-hole fluctuations.
That is, we subtract the correlation function obtained from the ground state with only a single dopant hole to remove fluctuation-fluctuation and fluctuation-dopant contributions and leave only the connected dopant-dopant correlations.
In Appendix~\ref{eq:corrected_corr}, we provide the details of this correction procedure and show the uncorrected correlation function, which is dominated by strong nearest-neighbor anticorrelation and a positive background signal.
The fact that spontaneously formed doublon-hole pairs cloud the signal in this way renders a quantitative analysis of the charge order significantly more challenging in the Fermi-Hubbard Hilbert space. \\
In the spin domain, the staggered correlation function
\begin{equation}
    C_{\hat{s}}(x, y) = 
    (-1)^{(x + y)} \langle \hat{s}^z_{1, y_0} \hat{s}^z_{x, y} \rangle
\end{equation}
with respect to a reference site on the edge of the cylinder shows an antiferromagnetic pattern extending over the entire system for the $t$-$J$ model.
However, a domain wall -- indicative of stripe formation -- is present in the Fermi-Hubbard ground state, marking a pronounced difference between the two models. The data for the $t$-$J$-3s model also shows this domain wall, suggesting the singlet-hopping term as the origin of this discrepancy.
%%%%%%%%%%%%%%%%%%%%%%%%%%%%%%%%%%
\section{Interpolation}
\label{sec:Interpolation}
To carry out a more quantitative analysis and to gain a detailed understanding of the changes in pair structure, we now consider the interpolating Hamiltonians $\hat{H}_{t \myphen J \myphen U}(\lambda)$ and $\hat{H}_{t \myphen J \myphen 3 \mathrm{s}}(\lambda)$.
As we already observed qualitatively, the pair becomes spatially more tightly bound when approaching the $t$-$J$ model at $\lambda \rightarrow 1$. This is confirmed quantitatively by a decreasing binding radius
\begin{equation}
    \langle \hat{r}^\mathrm{hh} \rangle =
    \frac{\sum_{\mathbf{i_0}} \sum_{\mathbf{j} \neq \mathbf{i_0}} \ |\mathbf{i_0} - \mathbf{j}| \ \langle \hat{n}^\mathrm{h}_\mathbf{i_0} \hat{n}^\mathrm{h}_\mathbf{j} \rangle}
    {\sum_{\mathbf{i_0}} \sum_{\mathbf{j} \neq \mathbf{i_0}} \ \langle \hat{n}^\mathrm{h}_\mathbf{i_0} \hat{n}^\mathrm{h}_\mathbf{j} \rangle} \ ,
    \label{eq:rhh}
\end{equation}
presented in Fig.~\ref{fig_lambda_scan}b.
For $\hat{H}(\lambda)_{t \myphen J \myphen U}$, the hole-hole correlation function is again corrected for doublon-hole fluctuations. To deal with the remaining background signal and allow for a meaningful comparison to $\hat{H}(\lambda)_{t \myphen J \myphen 3 \mathrm{s}}$, we limit the sums in~(\ref{eq:rhh}) to distances $| \mathbf{i_0} - \mathbf{j}| \leq 4$ for both models. \\
\\
A more accurate comparison of the interpolation Hamiltonians, unaffected by doublon-hole fluctuations, is afforded by the spin sector:
To quantify the appearance of the domain wall, Fig.~\ref{fig_lambda_scan}c illustrates the staggered spin-spin correlations across the full length of the cylinder.
This analysis shows that, when approaching the $t$-$J$ model, the spin domain wall disappears between $\lambda=0.6$ and $\lambda=0.8$.
Remarkably, the curves for $\hat{H}_{t \myphen J \myphen U}(\lambda)$ and $\hat{H}_\mathrm{t \myphen J \myphen 3 \mathrm{s}}(\lambda)$ coincide almost exactly, suggesting that the $t_3$ term captures the differences in pair structure between the Fermi-Hubbard and $t$-$J$ models in a quantitative way. \\
This finding is further supported by nearly identical energies per dopant
\begin{equation}
    \epsilon_2 = \frac{E_{2\mathrm{h}} - E_{0\mathrm{h}}}{2} \ ,
\end{equation}
comparing $\hat{H}(\lambda)_{t \myphen J \myphen U}$ and $\hat{H}(\lambda)_{t \myphen J \myphen 3 \mathrm{s}}$ (data presented in Appendix~\ref{app: Energy per Hole}).
%~\cite{SI}).
As shown in Fig.~\ref{fig_lambda_scan}d, the binding energy
\begin{equation}
    E_\mathrm{b} = (E_\mathrm{2h} - E_\mathrm{0h}) - 2 (E_\mathrm{1h} - E_\mathrm{0h}) \ ,
\end{equation}
which we obtain from separate ground-state searches in the sectors with $0$, $1$, and $2$ dopants, differs by more than a factor of $2$ between the Fermi-Hubbard and $t$-$J$-type models.
We find that the discrepancy is predominantly accounted for by the single-dopant energy $\epsilon_1 = E_\mathrm{1h} - E_\mathrm{0h}$. It is important to keep in mind, however, that $E_\mathrm{b} \ll \epsilon_{1; \: 2}$.
Interpolating between the $t$-$J$ and $t$-$J$-3s models, the binding energy is almost constant despite the changes in pair structure, suggesting that the two distinct pair configurations are similarly strongly bound.
The combination of these features supports an interpretation in terms of two coherently coupled, near-degenerate contributions to pairing -- mirroring the competition between stripe order and uniform superconductivity that characterizes the finite-doping phase diagrams.
We further motivate this interpretation in the language of a simple 2 level model in Appendix~\ref{app: two-level model}.
\\
To perform a quantitative analysis of the two contributions,
\begin{figure}[t!!]
\centering
\includegraphics[width=\linewidth]{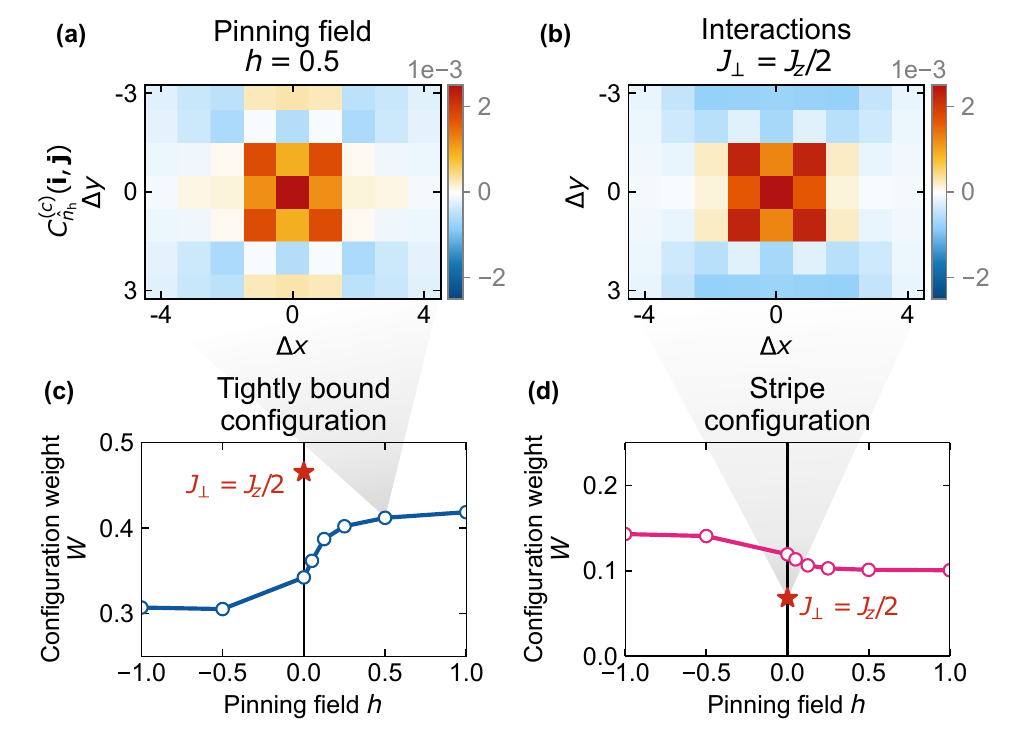}
\caption{
{Pinning field:}~Impacts of modifications in the spin sector on the charge order in the $t$-$J$-3s model. \textbf{(a, b)}~Hole-hole correlation functions: (a) An antiferromagnetic pinning field $h > 0$ on the edges of the system weakens the positive correlations in the stripe configuration compared to the case $h=0$ presented in Fig.~\ref{fig_overview_correlators}d. 
The nearest-neighbor and next-nearest-neighbor correlations, corresponding to tightly bound holes, stay large and positive.
(b) A modification to the spin interactions has a qualitatively similar but stronger effect. In this case, the stripe configuration is eliminated as the correlations at $\Delta y = 3$; $\Delta x \in \{-1, 0, 1 \}$ turn negative.
\textbf{(c, d)}~Quantitative impact of a pinning field of varying strength $h$ on the weights of the tightly bound (c) and stripe-like (d) charge configurations.
As observed in (a), turning on a pinning field $h > 0$ shifts weight from the stripe to the tightly bound configuration.
When pinning a domain wall in the system ($h < 0$), the effect is reversed.
The magnitudes of the changes are comparable to modifying the spin-interactions to $J_\perp = J/2$.
}
\label{fig_h_and_J_perp}
\end{figure}
\ignorespaces
we define the weights of each of the two charge configurations identified earlier as
\begin{equation}
    W =
    \frac{
    \sum\limits_{\mathbf{i_0}} \: \sum\limits_{\mathbf{\Delta i} \: \in \: \mathrm{config.}} \: \langle \hat{n}^\mathrm{h}_\mathbf{i_0} \hat{n}^\mathrm{h}_\mathbf{i_0 + \Delta i} \rangle}
    {\sum_{\mathbf{i_0}} \sum_{\mathbf{j} \neq \mathbf{i_0}} \ \langle \hat{n}^\mathrm{h}_\mathbf{i_0} \hat{n}^\mathrm{h}_\mathbf{j} \rangle} \ ,
\end{equation}
where $\mathbf{i_0}$ is again restricted to the center of the system. We define the tightly bound configuration as nearest-neighbor or next-nearest-neighbor pairs, i.e., in this case, the sum over $\Delta i = (\Delta x, \Delta y)$ runs over $|\Delta x| \leq 1$; $|\Delta y| \leq 1$. The stripe configuration is identified with $|\Delta x| \leq 1$; $|\Delta y| = L_y / 2 = 3$. \\
As showcased by Fig.~\ref{fig_configuration_weights}a, we find that the weight of the tightly bound pair (stripe) increases (decreases) monotonically with increasing $\lambda$, which is consistent with the singlet hopping term introducing a repulsion between the holes~\cite{Ammon1995-EffectThreesiteHopping}. \\
A more striking feature presents itself by investigating the tendency to form a spin domain wall separately for each charge configuration.
Introducing the $4$-point correlation function
\begin{align}
    C_{\hat{s}}(x, y | \mathbf{i}, \mathbf{j}) = 
    (-1)^{(x+y)}
    \frac{
    \langle
    \hat{n}^\mathrm{h}_\mathbf{i}
    \hat{n}^\mathrm{h}_\mathbf{j}
    \hat{s}^\mathrm{z}_{1, y_0}
    \hat{s}^\mathrm{z}_{x, y}
    \rangle
    }{
    \langle
    \hat{n}^\mathrm{h}_\mathbf{i}
    \hat{n}^\mathrm{h}_\mathbf{j}
    \rangle
    } \ ,
\end{align}
we find the correlator across the system to be negative for $\mathbf{i}$; $\mathbf{j}$ in the stripe-like configuration for all $\lambda$ and positive (except for $\lambda = 0$) in the tightly bound configuration, see Fig.~\ref{fig_configuration_weights}b.
The full spin-spin correlation maps conditioned on the hole locations (Figs.~\ref{fig_configuration_weights}c,d)
show the uninterrupted AFM pattern accompanying the tightly bound configuration and the spin domain wall in the stripe-like configuration.
We also confirm that the stripe-like charge configuration and the spin domain wall spatially coincide in Fig.~\ref{fig_configuration_weights}c and Appendix~\ref{subsec:4-point corrs} -- solidifying the picture of a bound, partially filled stripe.
This qualitative difference in the spin-sector between the two charge configurations makes the coexistence of these orderings in the system's ground state quite remarkable.
Moreover, the weighted average of the two contributions suffices to explain the appearance of a spin domain wall in the full system, observed in Fig.~\ref{fig_lambda_scan}c. 
\\
The local spin structure around the pairs of holes is presented in Fig.~\ref{fig_configuration_weights}e,f.
In agreement with early numerical studies of pairing~\cite{White1997-HolePairStructuresa}, the tightly bound pair is accompanied by a strong singlet on the diagonal of a $2 \times 2$ plaquette.
In contrast, the stripe configuration is accompanied by two singlets on a single rung of the cylinder.
The next-nearest-neighbor correlations across this rung are negative, indicating the formation of the spin domain wall.
%%%%%%%%%%%%%%%%%%%%%%%%%%%%%%%%%%
\section{Magnetic Pinning Field}
\label{sec:pinning_field}
To determine the hierarchy between charge and magnetic order, we study how modifications in the spin sector affect the pair structure.
Firstly, we introduce an AFM pinning field on the edges of the cylinder
\begin{equation}
    \hat{H}_h = 
    |h| \sum_{\substack{(x, y) \\ x \in \{ 1, L_x \}}} (-1)^{y} (- \mathrm{sgn}(h))^x \hat{s}^z_{(x, y)} \ .
\end{equation}
The sign convention is chosen such that $h > 0$ pins an uninterrupted AFM pattern, while $h < 0$ pins a domain wall.
Such a pinning field is often used in numerical studies to make spin-spin correlations accessible via local expectation values~\cite{Xu2024-CoexistenceSuperconductivityPartially, Shen2024-GroundStateElectrondopeda}.
To put the potential changes into perspective, we contrast the pinning field with the effects of modified spin interactions that also break the $\mathrm{SU}(2)$ symmetry.
That is, we compare to a case where we replace the Heisenberg term in $\hat{H}_{t \myphen J \myphen 3 \mathrm{s}}$ with XXZ interactions:
\begin{align}
     J \: \mathbf{\hat{S}}_{\mathbf{i}} \cdot \mathbf{\hat{S}}_{\mathbf{j}}
     \rightarrow
     \frac{J_\perp}{2}
     \left( \hat{s}^+_{\mathbf{i}} \hat{s}^{-}_{\mathbf{j}} + \hat{s}^{-}_{\mathbf{i}} \hat{s}^{+}_{\mathbf{j}} \right)
     + J_z
     \left( \hat{s}^z_{\mathbf{i}} \hat{s}^{z}_{\mathbf{j}} \right)  \ ,
\end{align}
and set $J_\perp = J/2 < J_z = J$.
By weakening the flip-flop interactions, the overall antiferromagnetic correlations $\langle \mathbf{\hat{S}}_\mathbf{i} \cdot \mathbf{\hat{S}}_\mathbf{j} \rangle$ are enhanced which we expect to disfavor the formation of a spin-domain wall. Therefore, we expect a qualitatively similar response to a pinning field $h > 0$. \\
The results are shown in Fig.~\ref{fig_h_and_J_perp}, confirming the one-to-one connection between the charge and spin sectors promoted earlier:
Pinning the uninterrupted AFM pattern, or weakening the interactions $J_\perp$ significantly reduces the weight of the striped charge configuration, while pinning a spin domain wall suppresses the tightly bound pair state.
This is in line with, and extends upon, the results of the previous section, where we find the different charge configurations to correlate with the respective spin states with and without a domain wall. \\
We point out that the magnitude of changes in the charge sector induced by the pinning field is comparable to those resulting from the modified interactions. This feature highlights the presence or absence of the domain wall (controlled via $h$) as the main driver of changes in the charge sector -- compared to other effects introduced by modifying the bulk microscopic interactions.
%%%%%%%%%%%%%%%%%%%%%%%%%%%%%%%%%%
\section{Discussion}
\begin{figure}[t!!]
\centering
\includegraphics[width=\linewidth]{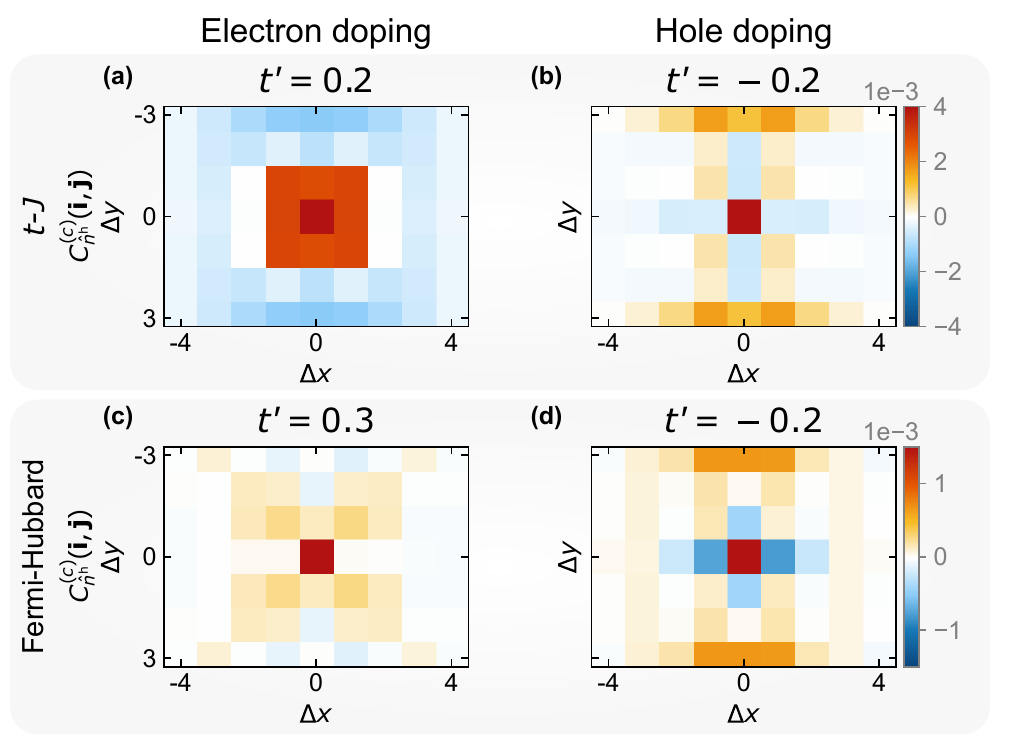}
\caption{
{Next-Nearest-Neighbor Hopping:}~Charge structure of the pair in the $t$-$J$~\textbf{(a, b)}~and Fermi-Hubbard~\textbf{(c, d)}~models in the presence of a next-nearest-neighbor tunneling term $t'$.
In both models, electron doping ($t' > 0$) (a, c) leads to tightly bound pairs, while hole doping ($t' < 0$) (b, d) has the dopants arranged in the stripe configuration.
}
\label{fig_t_prime}
\end{figure}

In summary, we find bound pairs with properties that we trace back to two coexisting contributions: a spatially tightly bound configuration and a stripe around the width $6$ cylinder, featuring a spin domain wall.
We argue that the contributions are near-degenerate, which is reflected in integer-pair striped and pair-density-wave phases reported in the finite-doping literature~\cite{Qin2020-AbsenceSuperconductivityPureb, Jiang2021-GroundstatePhaseDiagram, Jiang2024-GroundstatePhaseDiagram, Chen2023-DWavePairDensityWaveSuperconductivitya, Lu2024-EmergentSuperconductivityCompeting}. \\
While the pair structure is consistent across the Fermi-Hubbard, $t$-$J$-3s, and $t$-$J$ models, the relative weights of the two contributions are shifted when performing interpolations between these different Hamiltonians.
In line with arguments made for one-dimensional systems~\cite{Ammon1995-EffectThreesiteHopping}, the omission of the singlet-hopping term in the $t$-$J$ model leads to spatially more tightly bound pairs.
Its reintroduction quantitatively explains the emergence of a spin domain wall in the Fermi-Hubbard model. \\
As it mediates next-nearest-neighbor tunneling for the holes, the singlet hopping is intimately connected to the $t'$ term crucial to the physics of doped cuprates.
This term breaks the particle-hole symmetry and thus creates a distinction between the electron and hole-doped models.
In the cuprates, band-structure calculations estimate the strength of this term to be $t' \approx -0.2 \: t$~\cite{Andersen1995-LDAEnergyBands, Hirayama2018-InitioEffectiveHamiltonians}. Flipping the sign of $t'$ is equivalent to exchanging electron and hole dopants by means of a particle-hole transformation. \\
As an outlook, we show the pair structure in the presence of $t'$ in Fig.~\ref{fig_t_prime}.
%As expected, the effect is qualitatively similar to that of $t_3$ but
We find the term to shift the weight almost fully to either the tightly bound $(t' > 0)$ or stripe $(t' < 0)$ configuration.
This change in pair structure is in remarkable agreement with finite doping studies which report pairing and uniform-density superconductivity on the electron-doped $(t' > 0)$ side~\cite{Jiang2021-GroundstatePhaseDiagram, Chen2023-DWavePairDensityWaveSuperconductivitya, Lu2024-EmergentSuperconductivityCompeting, Xu2024-CoexistenceSuperconductivityPartially, Jiang2024-GroundstatePhaseDiagram} and find crystalline phases~\cite{Jiang2021-GroundstatePhaseDiagram, Jiang2024-GroundstatePhaseDiagram} and non-integer-pair stripes~\cite{Xu2024-CoexistenceSuperconductivityPartially} with increasing hole doping.
Here, the strong tendency to form spin-domain walls matches the rapid suppression of AFM order in the lightly hole-doped cuprates~\cite{Kastner1998-MagneticTransportOptical, Scalapino2012-CommonThreadPairing}.
\ignorespacesafterend
Notably, a superconductor formed from the tightly bound pairs we identify would be expected to behave BEC-like, which has been argued not to be the case for the cuprates~\cite{Sous2023-AbsenceBCSBECCrossover}.  
Thus, while numerical studies demonstrate superconducting correlations on the electron-doped side of the Fermi-Hubbard and $t$-$J$ models, we speculate that these models may feature a different kind of superconductivity compared to the materials they aim to describe, necessitating further studies. \\
Our finding that pairing and stripe formation are present on the single-pair level puts their competition within reach of simplified effective theories.
We believe that such a theory could be constructed by complementing our ground-state results with alternative approaches~\cite{Wang2015-OriginStrongDispersion, Carleo2017-SolvingQuantumManybodya, Bermes2024-MagneticPolaronsLinear} which can give access to pair spectra.
Additionally, quantum simulation experiments have emerged as a powerful tool for probing microscopic physics, having direct access to multi-point correlations at system sizes intractable by numerics.
In recent years, local charge and magnetic structures of dopants have been observed and interpreted in
terms of a geometric string picture~\cite{Chiu2019-StringPatternsDoped, Koepsell2021-MicroscopicEvolutionDopeda, Bohrdt2019-ClassifyingSnapshotsDoped} and efforts are ongoing to reach the low temperatures needed to observe pairing and eventually superconducting correlations.
In this regard, the binding energies presented in this work indicate that experimental setups working in the $t$-$J$ Hilbert space~\cite{Homeier2024-AntiferromagneticBosonicMathbit, Carroll2024-ObservationGeneralizedTJ} are at an advantage due to the absence of doublon-hole pairs.
%%%%%%%%%%%%%%%%%%%%%%%%%%%%%%%%%%
\section*{Data Availability}
The data presented in the figures of the main text is available at \url{https://github.com/TizianBlatz/pairing_structure_FH_tJ}.
\section*{Acknowledgments}
We are very grateful to Pit Bermes, Timothy J. Harris, Lukas Homeier, Christopher Roth, and Steven R. White for fruitful discussions and insights.
All DMRG calculations were performed using the SyTeN toolkit developed and maintained by C. Hubig, F. Lachenmaier, N.-O. Linden, T. Reinhard, L.
Stenzel, A. Swoboda, M. Grundner, S. Mardazad, F. Pauw, and
S. Paeckel. Information is available at \url{https://syten.eu}.
This research was funded by the Deutsche Forschungsgemeinschaft (DFG, German Research Foundation) under Germany’s Excellence Strategy—EXC-2111—390814868
% Fabian
and by the European Research Council (ERC) under the European Union’s Horizon 2020 research and innovation programme (grant agreement number 948141).
% KCS cluster
The work was supported by grant INST 86/1885-1 FUGG of the German Research Foundation (DFG).
\FloatBarrier
%\clearpage
\appendix
%%%%%%%%%% Merge with methods %%%%%%%%%%
%%%%%%%%%% Prefix a "M" to all equations, figures, and tables and reset the counter %%%%%%%%%%
%\setcounter{equation}{0}
%\setcounter{figure}{0}
%\setcounter{table}{0}
%\setcounter{page}{1}
%\makeatletter
%\renewcommand{\theequation}{S\arabic{equation}}
%\renewcommand{\thefigure}{S\arabic{figure}}
%\textbf{\large Supplementary Materials} \\
%%%%%%%%%%%%%%%%%%%%%%%%%%%%%%%%%%
\section{Doublon-Hole Correction}
\begin{figure}[t!!]
\centering
\includegraphics[width=\linewidth]{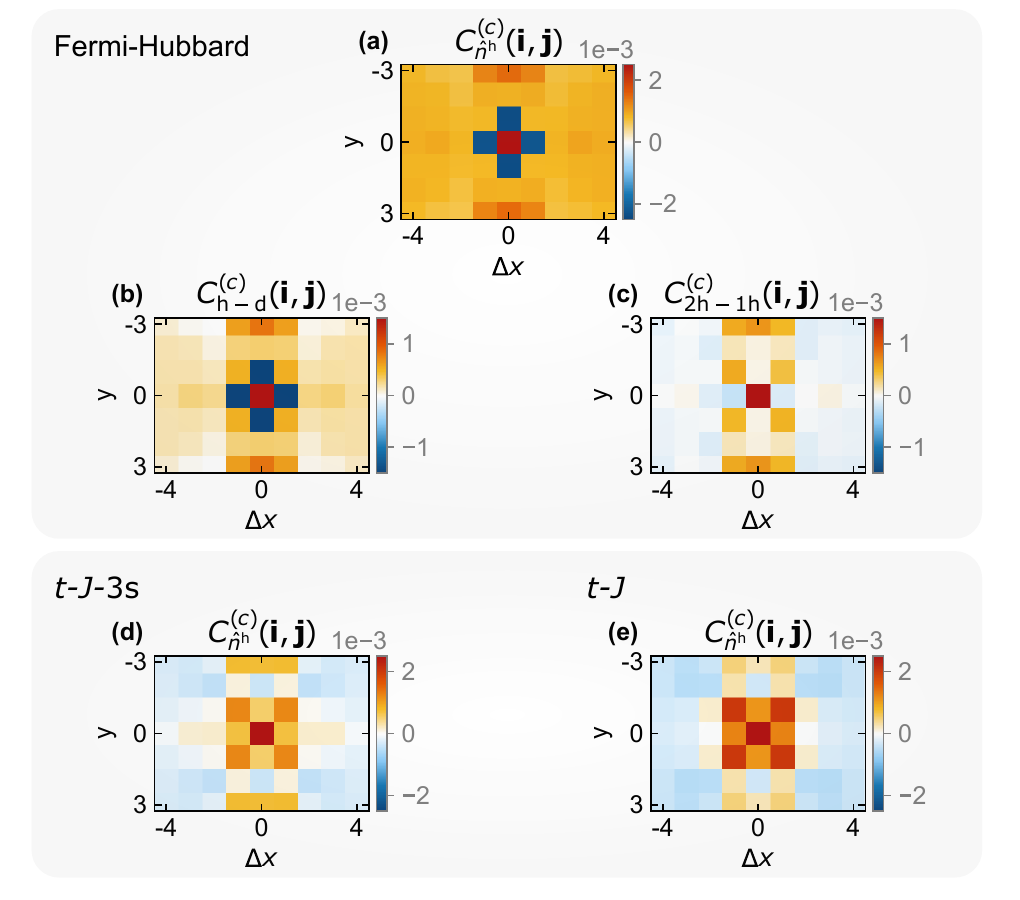}
\caption{
{Doublon Correction:} Comparison of the pair structure of doped charge carriers for the Fermi-Hubbard and $t$-$J$-type models. The color coding truncates the large positive signal in the center, as well as the strongly negative nearest-neighbor doublon-hole signal in the Fermi-Hubbard model.
\textbf{(a)} Connected correlations function $\langle \hat{n}^\mathrm{h} \hat{n}^\mathrm{h} \rangle^\mathrm{(c)} = \langle \hat{n}^\mathrm{h} \hat{n}^\mathrm{h} \rangle - \langle \hat{n}^\mathrm{h} \rangle \langle \hat{n}^\mathrm{h} \rangle$ for the Fermi-Hubbard model as presented in Fig.~\ref{fig_overview_correlators}.
\textbf{(b, c)} Approaches to modify (a) to correct for doublon-hole fluctuations. (b)~Subtracting the doublon-doublon correlator as a proxy for correlations between non-dopant holes.
(c)~Subtracting the correlations obtained from a calculation with a single dopant hole. This approach uncovers the clearest signature in our numerical study but is not immediately available for experiments.
~\textbf{(d, e)} Connected correlation functions for the $t$-$J$-type models as presented in Fig.~\ref{fig_overview_correlators}.
}
\label{supp_fig_doublon_correction}
\end{figure}

Without corrections, the hole-hole correlation function in the Fermi-Hubbard model is dominated by nearest-neighbor anticorrelation and a positive background signal (see Fig.~\ref{supp_fig_doublon_correction}) -- obscuring the pair structure of dopant charge carriers and complicating a quantitative analysis.
We attribute this feature to the presence of doublon-hole fluctuations in the Fermi-Hubbard model.
Despite the large coupling $U/t=8$ employed here, the number of holes produced in this way is significantly larger than the maximum number of $2$ dopant holes, leading to the strongly anticorrelated nearest-neighbor signal dominating the hole-hole correlation function. \\
Here, we present two ways to correct the fluctuation effects to uncover the pair structure of the dopant holes, similar to the $t$-$J$-3s model: \\
The most natural approach is to subtract doublon-doublon correlations from the hole-hole correlation function, i.e.,
\begin{align}
    C^{(\mathrm{c})}_{\mathrm{h - d}}(\mathbf{i}, \mathbf{j}) = C^{(\mathrm{c})}_{\hat{n}^\mathrm{h}}(\mathbf{i}, \mathbf{j}) - C^{(\mathrm{c})}_{\hat{n}^\mathrm{d}}(\mathbf{i}, \mathbf{j}) \ ,
\end{align}
where $C^{(c)}_{\hat{n}^\mathrm{h}}(\mathbf{i}, \mathbf{j})$ is defined in (\ref{eq:connected_corr}) and $C^{(c)}_{\hat{n}^\mathrm{h}}(\mathbf{i}, \mathbf{j})$ replaces the hole number operator $\hat{n}^\mathrm{h}$ for the corresponding operator $\hat{n}^\mathrm{d}$ counting double occupancies.
This amounts to subtracting fluctuation-fluctuation correlations that do not involve dopant holes. Notably, this type of correction is available to quantum simulation experiments (as long as the employed detection scheme can differentiate double occupancies from holes~\cite{Boll2016-SpinDensityresolvedMicroscopy, Koepsell2020-RobustBilayerCharge, Hartke2020-DoublonHoleCorrelationsFluctuation}). \\
A more sophisticated type of correction is afforded by precise control of the doping level:
By subtracting the doublon corrected hole-hole correlator obtained from calculations with a single dopant hole from that obtained from a pair of dopants, we remove all parts of the correlation function involving holes originating from fluctuations.
That includes dopant-fluctuation contributions which are not corrected for by the first approach.
The corrected correlation function is defined as
\begin{align}
    C^{(\mathrm{c})}_{\mathrm{2h - 1h}}(\mathbf{i}, \mathbf{j}) = C^{(\mathrm{c}), \: (2)}_{\mathrm{h - d}}(\mathbf{i}, \mathbf{j}) - 2 \frac{N^\mathrm{d}_{(2)}}{N^\mathrm{d}_{(1)}} \: C^{(\mathrm{c}), \: (1)}_{\mathrm{h - d}}(\mathbf{i}, \mathbf{j}) \ ,
    \label{eq:corrected_corr}
\end{align}
where the number in brackets refers to the number of dopants $L - N$ and the normalization factor accounts for the different number of dopants and expected number of holes from fluctuations between the two calculations. This definition of the correlation function is used to obtain the pair structure in the Fermi-Hubbard model presented in the main text.
The binding radius presented in Fig.~\ref{fig_lambda_scan}b is evaluated according to the unconnected version of (\ref{eq:corrected_corr}), i.e., without subtracting density terms.
\\
The correlation functions obtained from correcting the Fermi-Hubbard data are compared to those presented for the $t$-$J$-type models in Fig.~\ref{supp_fig_doublon_correction}.
The first correction approach succeeds in removing most of the positive background signal observed at longer ranges and reveals slightly enhanced next-nearest-neighbor correlations.
The more sophisticated correction approach makes this trend even clearer and removes most of the strong negative nearest-neighbor signal.
The positive correlations observed for the stripe configuration and on next-nearest-neighbor sites are about equal in strength -- consistent with the $t$-$J$-3s model. \\
This adds further support to our interpretation of the $\lambda$-scan, where we argue that the discrepancies between the Fermi-Hubbard and $t$-$J$ models are largely explained by the $t_3$ term.
For a quantitative comparison of the weights of different charge configurations -- as presented in Figures~\ref{fig_configuration_weights} and \ref{fig_h_and_J_perp} of the main text -- it remains beneficial to work in the $t$-$J$ Hilbert space.
%%%%%%%%%%%%%%%%%%%%%%%%%%%%%%%%%%
\section{Normalization Factor in $C^{(c)}_{\hat{n}^\mathrm{h}}(\mathbf{i}, \mathbf{j})$}
In the thermodynamic limit, the connected density-density correlation function is given by
\begin{align}
   \langle \hat{n}^\mathrm{h}_\mathbf{i} \hat{n}^\mathrm{h}_\mathbf{j} \rangle - \langle \hat{n}^\mathrm{h}_\mathbf{i} \rangle \langle \hat{n}^\mathrm{h}_\mathbf{j} \rangle \ . 
\end{align}
However, at low particle numbers, combinatoric effects begin to play an increasingly important role, and a normalization factor is introduced to ensure that the connected correlations vanish for an uncorrelated state. For such a state, the $n$-point correlation function is given by the number of possibilities to choose $n$ holes from the total number $N^\mathrm{h}$, divided by the number of choices for $n$ lattice sites from the total number $L$. This gives the $2$-point correlator of randomly placed holes as
\begin{align}
    \langle \hat{n}^\mathrm{h}_\mathbf{i} \hat{n}^\mathrm{h}_\mathbf{j} \rangle_\mathrm{random}
    = \frac{N^\mathrm{h} (N^\mathrm{h} - 1)}{L (L - 1)} \ .
\end{align}
The respective density term is given by
\begin{align}
    \langle \hat{n}^\mathrm{h}_\mathbf{i} \rangle \langle \hat{n}^\mathrm{h}_\mathbf{j} \rangle_\mathrm{random}
    = \frac{(N^\mathrm{h})^2}{L^2} \ ,
\end{align}
leading to a normalization factor
\begin{align}
    \frac{N^\mathrm{h} - 1}{N^\mathrm{h}} \frac{L}{L - 1} \ .
\label{eq:normalization factor}
\end{align}
Due to the large lattice size, we neglect the $L$-dependent part of (\ref{eq:normalization factor}), equivalent to reducing the factor to that of particles without a hardcore constraint (see e.g., \cite{Leonard2023-RealizationFractionalQuantum}). In the Fermi-Hubbard Hilbert space, where $N^\mathrm{h}$ is not a good quantum number, we replace it by $\langle N^\mathrm{h} \rangle$, arriving at (\ref{eq:connected_corr}). 
%%%%%%%%%%%%%%%%%%%%%%%%%%%%%%%%%%
\section{DMRG}
We use DMRG in the framework of matrix product states (MPS)~\cite{Schollwoeck2011-DensitymatrixRenormalizationGroup} to calculate the ground state of up to two dopants with respect to half-filling, i.e., for a lattice with $L$ sites, we consider the  $L - 2 \leq N \leq L$ particle sectors.
The calculations are performed on cylindrical systems where we define the y-direction as going around the cylinder (periodic boundary conditions) while the x-direction is defined parallel to the cylinder axis (open boundary conditions). Throughout this work, we consider $6$-legged cylinders with system size $L_y \times L_x = 6 \times 12$. \\
Since we consider only a single pair of holes -- not a fixed doping level -- the finite system size is crucial for learning about the system's tendencies to form stripes at finite doping.
If bound, the holes may form a stripe-like structure with non-vanishing filling $1/3$.
\\
The DMRG simulations are performed using the SyTeN toolkit.
Depending on the model, we work in either the Fermi-Hubbard ($\hat{H}_\mathrm{FH}, \ \hat{H}_{t \myphen J \myphen U}(\lambda)$), or $t$-$J$ ($\hat{H}_{t \myphen J \myphen 3 \mathrm{s}}(\lambda)$) Hilbert space. Whenever the $\mathrm{SU}(2)$ spin symmetry of the models is not broken, we exploit the full $\mathrm{U}(1) \times \mathrm{SU}(2)$ symmetry of the particle number and spin, working in the ($N = L - 2, \ S = 0$) sector. If the $\mathrm{SU}(2)$ symmetry is broken e.g., by pinning fields on the edges of the system, calculations are instead performed in the ($N = L - 2, \ S^z = 0$) sector of the reduced $\mathrm{U}(1) \times \mathrm{U}(1)$ symmetry.
To present observables uniformly throughout this work, we always provide spin-information in the $\hat{s}^z$ basis – directly accessible in $\mathrm{U}(1)$ calculations, or determined via $\langle \hat{s}^z_\mathbf{i} \hat{s}^z_\mathbf{j} \rangle = 1/3 \: \langle \mathbf{\hat{S}}_\mathbf{i} \cdot \mathbf{\hat{S}}_\mathbf{j} \rangle$ when the $\mathrm{SU}(2)$ symmetry is preserved. \\
For the interpolation Hamiltonian $\hat{H}(\lambda)_{t \myphen J \myphen U}$, we perform calculations in the Fermi-Hubbard Hilbert space up to a maximum value of $\lambda = 0.9$. For $\lambda = 1$ double occupancies suffer an infinite energy penalty, so the calculation is performed in the $t$-$J$ Hilbert space.\\
To find the ground state, we use a mixture of the single-site and two-site DMRG algorithms. The observables we investigate are well-converged using bond dimensions up to $m = 6144$, where we find the effective $\mathrm{U}(1)$ bond dimension of $\mathrm{SU}(2)$-calculations with $m \approx 6 \, 000$ to be $m^\mathrm{eff.} \approx 18 \, 000$.
%\FloatBarrier
%\clearpage
%\appendix
%%%%%%%%%% Merge with supplemental materials %%%%%%%%%%
%%%%%%%%%% Prefix a "S" to all equations, figures, and tables and reset the counter %%%%%%%%%%
%\setcounter{equation}{0}
%\setcounter{figure}{0}
%\setcounter{table}{0}
%%\setcounter{page}{1}
%\makeatletter
%\renewcommand{\theequation}{S\arabic{equation}}
%\renewcommand{\thefigure}{S\arabic{figure}}
%\textbf{\large Supplemental Materials} \\
%\\
\begin{figure}[t!!]
\centering
\includegraphics[width=\linewidth]{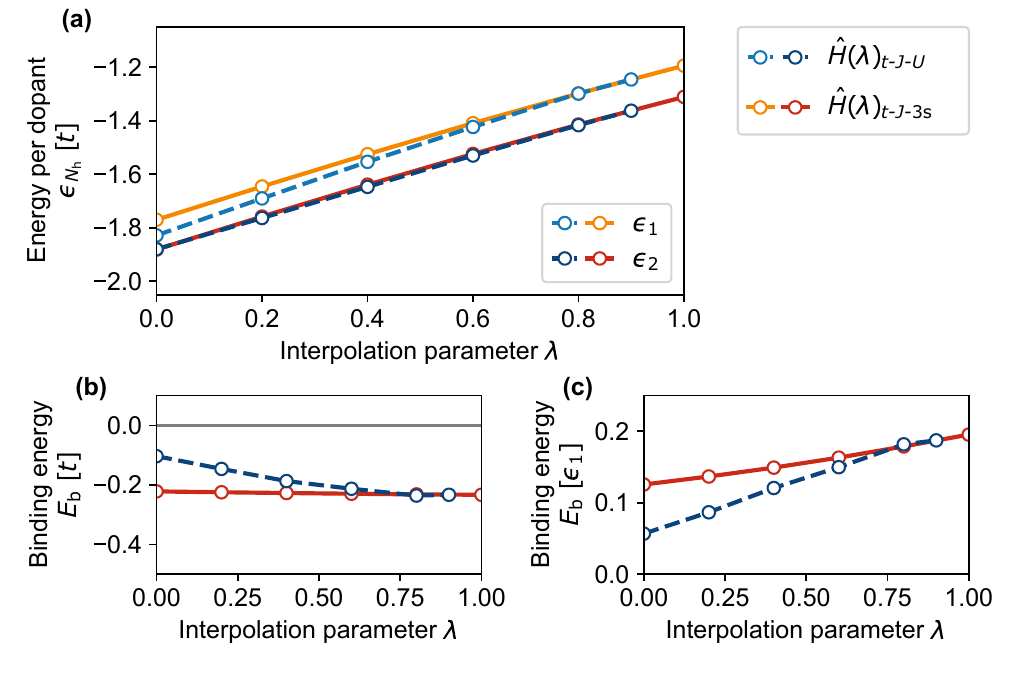}
\caption{
{Energies:}~\textbf{(a)} Energies per hole $\epsilon_{1; \: 2}$ from calculations with 1 and 2 dopant holes. Comparing $\hat{H}(\lambda)_{t \myphen J \myphen U}$ and $\hat{H}(\lambda)_{t \myphen J \myphen 3 \mathrm{s}}$, the curves for $\epsilon_2$ coincide. The values $\epsilon_1$ differ, as the Hamiltonians $\hat{H}_\mathrm{FH}$ and $\hat{H}_{t \myphen J}$ are approached in the $\lambda \rightarrow 0$ limit.
\textbf{(b, c)} Binding energy $E_\mathrm{b}$ calculated from the difference between $\epsilon_2$ and $\epsilon_1$ in units of $t$ (b) and $\epsilon_1$ (c). 
}
\label{supp_fig_e_per_hole}
\end{figure}
\section{Binding Energy and Energy per Hole}
\label{app: Energy per Hole}
In Fig.~\ref{fig_lambda_scan}d of the main text, we present the binding energy $E_\mathrm{b}$ of a pair of dopants which is calculated from the energy per dopant
\begin{align}
        \epsilon_{N_\mathrm{h}} = \frac{E_{N_\mathrm{h}} - E_{0\mathrm{h}}}{N_\mathrm{h}} \\
        \mathrm{as} \quad
        E_\mathrm{b} = 2 (\epsilon_2 - \epsilon_1) \ .
\end{align}
Here, we investigate the components $\epsilon_{1; \: 2}$ separately to determine the origin of the smaller binding energy in the Fermi-Hubbard model compared to $\hat{H}_{t \myphen J \myphen 3 \mathrm{s}}$. \\
As displayed in Fig.~\ref{supp_fig_e_per_hole}, $\epsilon_2$ is almost identical for $\hat{H}(\lambda)_{t \myphen J \myphen U}$ and $\hat{H}(\lambda)_{t \myphen J \myphen 3 \mathrm{s}}$ throughout the interpolation. In particular, the large difference $\epsilon^{\mathrm{(FH)}}_2 - \epsilon^{(t \myphen J)}_2 \approx 0.57 \: t$ is accounted for within the accuracy of our numerics by introducing the $t_3$ term, complementing the closely matched pair properties we observe between $\hat{H}_\mathrm{FH}$ and $\hat{H}_{t \myphen J \myphen 3 \mathrm{s}}$ in the main part of our work. \\
Consequently, the difference in binding energy between these two models stems almost exclusively from the energy of a single dopant.
As we note in the main text, $| E_\mathrm{b}^{t \myphen J} | \approx | E_\mathrm{b}^{t \myphen J \myphen 3 \mathrm{s}} | \approx 0.23 \: t$ does not change significantly with $\lambda$ for the $t$-$J$-type models.
This value agrees well with results for open boundary conditions~\cite{White1997-HolePairStructuresa}, indicating that the $6$-leg geometry is
free of the strong dependence on boundary conditions established for width $4$ systems~\cite{Chung2020-PlaquetteOrdinaryWave}. \\
In contrast, we find a notably lower value $| E_\mathrm{b}^{(\mathrm{FH})} | \approx 0.1 \: t$ in the Fermi-Hubbard model, which will likely translate to lower temperatures required to observe pairing effects experimentally.
\begin{figure}[t!!]
\centering
\includegraphics[width=\linewidth]{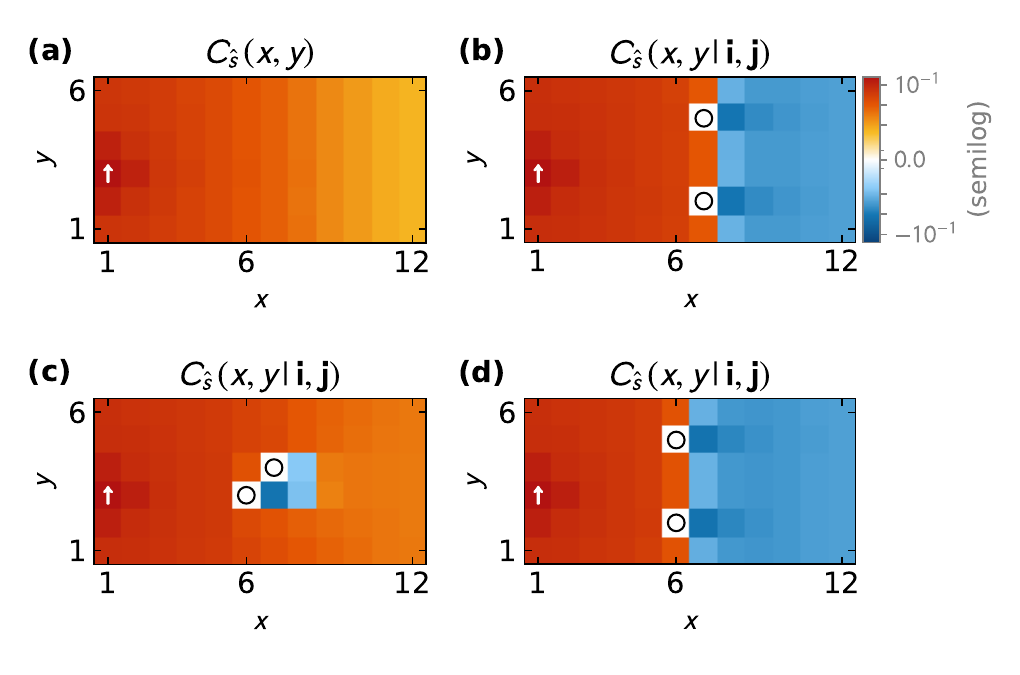}
\caption{
{Four Point Correlations:}~Staggered spin-spin correlation functions for different locations of the dopant charge carriers.~\textbf{(a)} Two-point correlation function $C_{\hat{s}} \: (x, y)$ for the $t$-$J$ model, as shown in Fig.~\ref{fig_overview_correlators}h). The location of the reference spin is indicated by a white arrow.
\textbf{(b, c, d)} Four-point correlation function $C_{\hat{s}} \: (x, y \: | \: \mathbf{i}, \mathbf{j})$, providing the spin-spin correlations with the positions of the two dopants fixed.
The fixed hole positions are indicated by black circles.
As established in Fig.~\ref{fig_configuration_weights}, the tightly bound configuration of dopants (c) is accompanied by uninterrupted AFM order, while the stripe-like configuration (b, d) shows a spin domain wall.
The location of the spin-domain wall coincides with that of the holes, which we confirm by shifting the charges to the left by one site in (d).
}
\label{supp_fig_domain_wall}
\end{figure}
\section{Spatially Resolving the Domain Wall}
\label{subsec:4-point corrs}
In the main text, we have established a correspondence between the stripe-like charge configuration and the presence of a domain wall in the spin sector.
This one-to-one correspondence is further supported by a detailed analysis of the spin sector through the four-point correlation functions defined in Section~\ref{sec:Interpolation}.
Fig.~\ref{supp_fig_domain_wall} displays the AFM domains over the full lattice for fixed dopant positions. Making use of this full spatial resolution, we can confirm that the location of the spin-domain wall exactly coincides with that of the stripe-like configuration of charges.
\section{Two-Level Model}
\label{app: two-level model}
In the main text, we argue that we can think of the tightly-bound and the stripe-like configurations as coherently coupled, near degenerate contributions to pairing.
Here, we provide further motivation for this picture by comparing our results to the expectations from a simple, phenomenological two-level model:
We assume that we can write the ground state of the system as a superposition of the two configurations:
\begin{equation}
    \ket{ \mathrm{GS}} = \alpha \: \ket{\textrm{tightly-bound}} + \beta \: \ket{\textrm{stripe-like}} \ .
\end{equation}
We take the two contributions to be orthonormal, so the Schrödinger-equation $\hat{H} \ket{ \mathrm{GS}} = E_0 \ket{ \mathrm{GS}}$ in the basis of the two configurations becomes
%\begin{equation}
%    \hat{H} \ket{ \mathrm{GS}}
%    = \begin{bmatrix} E_\alpha & \Delta \\ \Delta^* & E_\beta \end{bmatrix}
%    \begin{bmatrix} \alpha \\ \beta \end{bmatrix}
%    = E_0 \begin{bmatrix} \alpha \\ \beta \end{bmatrix} \ .
%\end{equation}
\begin{equation}
    \hat{H} \ket{ \mathrm{GS}}
    = \begin{bmatrix} 0 & g \\ g & \Delta \end{bmatrix}
    \begin{bmatrix} \alpha \\ \beta \end{bmatrix}
    = E_0 \begin{bmatrix} \alpha \\ \beta \end{bmatrix} \ .
\end{equation}
Thereby, we have introduced two phenomenological parameters: the coupling $g$ and the detuning $\Delta$, which determine the ratio
\begin{equation}
    \frac{\alpha}{\beta} = \frac{1}{2 g} \left( \Delta - (\Delta^2 + 4 g^2)^{1/2} \right)
\end{equation}
in the ground state. The smooth control and wide range of these relative weights between the tightly-bound and stripe-like configuration observed in scans of the parameters $\lambda$, $h$, and $t'$ leads us to conclude that $g$ and $\Delta$ must be of similar magnitude.
That is, both coupling and a comparably small detuning are important to explain our observations. \\
As a concrete example, let us consider the scan of the pinning field $h$ presented in Section~\ref{sec:pinning_field}: This field introduces an energy penalty or gain for the spin domain wall present in the stripe-like configuration. Thus, in the two-level model we would expect $\Delta$ to increase with $h$.
As the interactions are not modified, the coupling $g$ is expected to change only little.
Taking the configuration weights $W$ in the charge sector as proxies for $\alpha^2 / \beta^2$, we can read off a shift from $\alpha^2 / \beta^2 \sim 1/2$ at $h = - t$ to $\alpha^2 / \beta^2 \sim 1/4$ at $h = \: t$ from Fig.~\ref{fig_h_and_J_perp}c,d.
In the two-level model, this translates to a change from $\Delta / g \sim 0.35$ to $\Delta / g \sim 0.75$ -- matching the anticipated increase with $h$ and similar magnitudes of $\Delta$ and $g$.
%\begin{align}
%    E_\pm &= 1/2 \left( \Delta \pm (\Delta^2 + 4 g^2)^{1/2} \right) \\
%    \Delta E^2 &= \Delta^2 + 4 g^2 \\
%    \alpha / \beta &= 1/2 g \left( \Delta \pm (\Delta^2 + 4 g^2)^{1/2} \right)
%\end{align}
\begin{figure}[t!!]
\centering
\includegraphics[width=\linewidth]{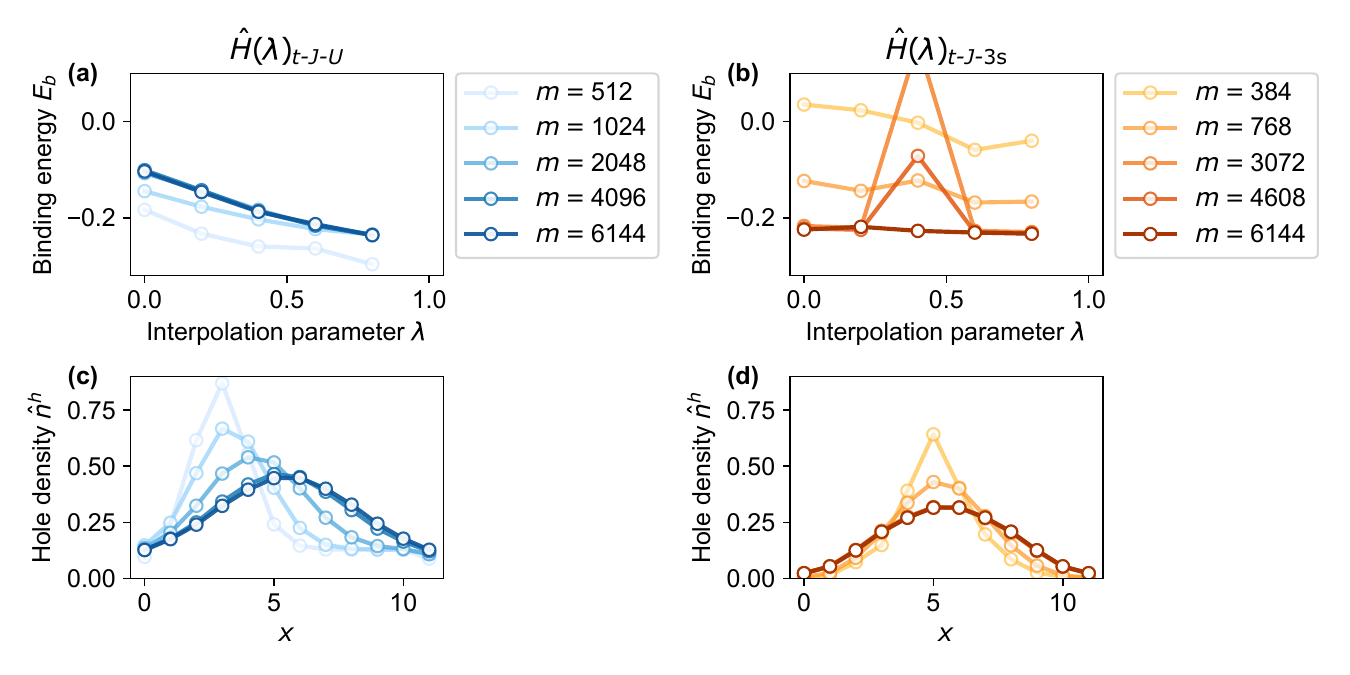}
\caption{
{DMRG Convergence:}~\textbf{(a, b)} Binding energy $E_\mathrm{b}$ as a function of the tuning parameter $\lambda$ and the bond dimension $m$ for $\hat{H}_{t \myphen J \myphen U}(\lambda)$ (a) and $\hat{H}_{t \myphen J \myphen 3 \mathrm{s}}(\lambda)$ (b). At low $m$, the calculation in the Fermi-Hubbard Hilbert space tends to overestimate the binding energy,  while $E_\mathrm{b}$ is underestimated in the $t$-$J$ Hilbert space. For $\lambda \rightarrow 1$, the values converge toward one another, which we use as a crosscheck between the different Hilbert spaces.
\textbf{(c, d)} Hole-densities $\hat{n}^\mathrm{h}$ a function of $m$ and coordinate $x$, summed over the periodic ($y$-) direction. Here, $\lambda = 0.0$ is fixed, i.e. we compare $\hat{H}_\mathrm{FH}$ to $\hat{H}_{t \myphen J \myphen 3 \mathrm{s}}$. The total number of holes $\langle \hat{N}^\mathrm{h} \rangle$ is larger in the Fermi-Hubbard Hilbert space (c) due to doublon-hole fluctuations. Convergence with $m$ is slower in the Fermi-Hubbard than in the $t$-$J$ Hilbert space.
The asymmetry at low $m$ corresponds to the direction of the first DMRG sweep; the initial state features uniform hole density.
}
\label{supp_fig_convergence}
\end{figure}
\section{DMRG Convergence}
To monitor the convergence of our calculations, we track the binding energy $E_\mathrm{b}$ and hole-densities $\langle n^h_\mathbf{i} \rangle$ with increasing bond dimension $m$.
As a small difference of energies -- with $|E_\mathrm{b}|/|E_0| \approx 0.01$ (and $E_0$ a typical ground-state energy) -- the binding energy is highly sensitive to the overall convergence, and in particular to the relative convergence of calculations with different numbers of dopants. As
\begin{align}
    \hat{H}_{t \myphen J \myphen U}(\lambda = 1) = \hat{H}_{t \myphen J \myphen 3 \mathrm{s}}(\lambda = 1) = \hat{H}_{t \myphen J} \ ,
\end{align}
we can crosscheck our calculations in the different Hilbert spaces.
We find this check to be highly valuable due to the distinct and complementary computational challenges faced in either Hilbert space at low doping.
While calculations in the larger Fermi-Hubbard Hilbert space require a significantly larger bond dimension to converge,
the calculation is less prone to getting stuck during sweeping as the motion of particles is much less constrained.
To reduce the risk of getting stuck, we utilize global subspace expansion, as proposed for use in time evolution by M. Yang and S. R. White~\cite{Yang2020-TimeDependentVariational}, to increase the bond dimension in the early stages of our calculations. \\
The convergence of $E_\mathrm{b}$ with the bond dimension is showcased in Fig.~\ref{supp_fig_convergence}.
The aforementioned differences in convergence are clearly visible, but the values of $E_\mathrm{b}$ approach one another as $\lambda \rightarrow 1$. \\
Complementary to the binding energy, we also present convergence data for the hole density at fixed $\lambda = 0.0$.
As changes of delocalization of two holes cost very little energy (a fraction of $t$), the hole density along the $x$-direction serves as a sensitive measure of convergence that is (in contrast to $E_\mathrm{b}$) mostly independent of the pairing properties. \\
Based on these properties, we can report good convergence for computations that are significantly less demanding than their finite doping counterparts. There, studies routinely find a sensitive dependence of the type of order stabilized on the initial state used in the DMRG~\cite{Jiang2021-GroundstatePhaseDiagram, Shen2024-GroundStateElectrondopeda}.
Comparing an antiferromagnetic Néel product state with localized dopants to a Fermi-sea state, we observe no such dependence.
While the same pairing order is stabilized, convergence is typically slower when starting from the localized state.
Hence, all data presented in this work is obtained from the more efficient Fermi-sea initial state.
%
%\bibliography{pairing_paper, suppmat}
%\bibliographystyle{apsrev4-2}

%
\end{document}